\documentclass[12pt,preprint]{aastex}
\begin{document}
\newcommand{\multi}{{\sc MultiNest}}
\newcommand{\cofiam}{{\tt CoFiAM}}
\title{KOI-142, THE KING OF TRANSIT VARIATIONS, IS A PAIR \\OF PLANETS NEAR THE 2:1 RESONANCE}
\author{David Nesvorn\'y$^1$, David Kipping$^2$, Dirk Terrell$^1$, Joel Hartman$^{3}$,\\
G\'asp\'ar \'A. Bakos$^{3}$, Lars A. Buchhave$^{4,5}$}
\affil{(1) Department of Space Studies, Southwest Research Institute, Boulder, CO~80302, USA} 
\affil{(2) Harvard-Smithsonian Center for Astrophysics, Cambridge, MA~02138, USA}
\affil{(3) Department of Astrophysical Sciences, Princeton University, Princeton, NJ~05844, USA}
\affil{(4) Niels Bohr Institute, University of Copenhagen, DK-2100, Copenhagen, Denmark}
\affil{(5) Natural History Museum of Denmark, University of Copenhagen, DK-1350, Copenhagen, Denmark}
\begin{abstract}
The Transit Timing Variations (TTVs) can be used as a diagnostic of gravitational interactions
between planets in a multi-planet system. Many Kepler Objects of Interest (KOIs) exhibit significant
TTVs, but KOI-142.01 stands out among them with an unrivaled, $\simeq$12-hour TTV amplitude.
Here we report a thorough analysis of KOI-142.01's transits. We discover periodic Transit Duration 
Variations (TDVs) of KOI-142.01 that are nearly in phase with the observed TTVs. We show that 
KOI-142.01's TTVs and TDVs uniquely detect a non-transiting companion with a mass $\simeq$0.7 
that of Jupiter (KOI-142c). KOI-142.01's mass inferred from the transit variations is consistent 
with the measured transit depth, suggesting a Neptune class planet (KOI-142b). The orbital 
period ratio $P_c/P_b=2.03$ indicates that the two planets are just wide of the 2:1 resonance. 
The present dynamics of this system, characterized here in detail, can be used to test various 
formation theories that have been proposed to explain the near-resonant pairs of exoplanets.
\end{abstract}
\section{Introduction}
The two methods used so far to discover the majority of exoplanets, transit observations (TOs) and radial 
velocity (RV) measurements, have known limitations. The best current RV sensitivity of $\sim$1 m s$^{-1}$ 
allows us to detect planets down to $\sim$1 Neptune mass at 1~AU, assuming a bright, solar-mass host star and 
$\sin I \sim 1$, where $I$ is the inclination of planet's orbit relative to the sky plane. The TO method, 
on the other hand, can be used to detect smaller planets (Borucki et al. 2011), but requires that $\sin 
I \simeq 1$. Most multi-planet systems discovered from TOs are therefore implied to be coplanar to within
$\simeq$1$^\circ$ (Lissauer et al. 2011a). 

Ideally, we would like to use the detection statistics of planets and planetary systems from TOs and RVs to 
build a population model that globally describes the distribution of planet properties (such 
as the planet occurrence depending on the host star mass and metalicity, planetary mass and size distributions, 
multiplicity, distributions of orbital spacing, eccentricity and inclination, etc.) in the Galaxy. 
Much of this information, however, is difficult to obtain from TOs and RV measurements alone. The problem is 
rooted in the fact that these methods are blind to certain classes of planets and planetary systems, and 
often allow us to obtain only approximate properties of the detected systems.

The Transit Timing Variation (TTV) method can help resolve some of these issues. The TTVs occur when 
gravitational perturbations produce deviations from strictly Keplerian orbits, such that the spacing of individual 
transits is not exactly periodic (Miralda-Escud\'e 2002, Agol et al. 2005, Holman \& Murray 2005). 
The TTVs have been used to confirm some of the 
transiting planet candidates from Kepler (e.g., Holman et al. 2010, Lissauer et al. 2011b), detect and characterize 
non-transiting planets (Nesvorn\'y et al. 2012), and search for moons (Kipping et al. 2012, 2013). The TTV 
analysis often provides specific information about the system that is unavailable from TOs or RVs alone. 

Here we discuss the Kepler Object of Interest (KOI) 142.01. We show that the observed TTVs can uniquely be fit
by a sub-Jovian mass planet near the exterior 2:1 orbital resonance with the transiting Neptune-class planet.
The Transit Duration Variations (TDVs) produced by the interaction of two planets are also detected. We use 
the TTVs and TDVs to investigate the near-resonant dynamics of the KOI-142 system. 
The methods and results are described in Sections 2 and 3. The broader implications of our work are discussed 
in Section 4. Efforts such as these can can help us to understand the formation of the near-resonant planetary 
systems. They can also provide important guidance in our pursuit of the planet population model. 
\section{Method}
The TTVs of KOI-142.01 were identified by Ford et al. (2011, 2012) and Steffen et al. (2012). 
Mazeh et al. (2013) updated KOI-142.01's TTVs using 
the first twelve quarters of the Kepler data, highlighted the large TTV amplitude ($\simeq$12 hours), and 
suggested that this `king of TTVs' can host one or more additional planets. 

\begin{figure}[t]
\epsscale{0.7}
\plotone{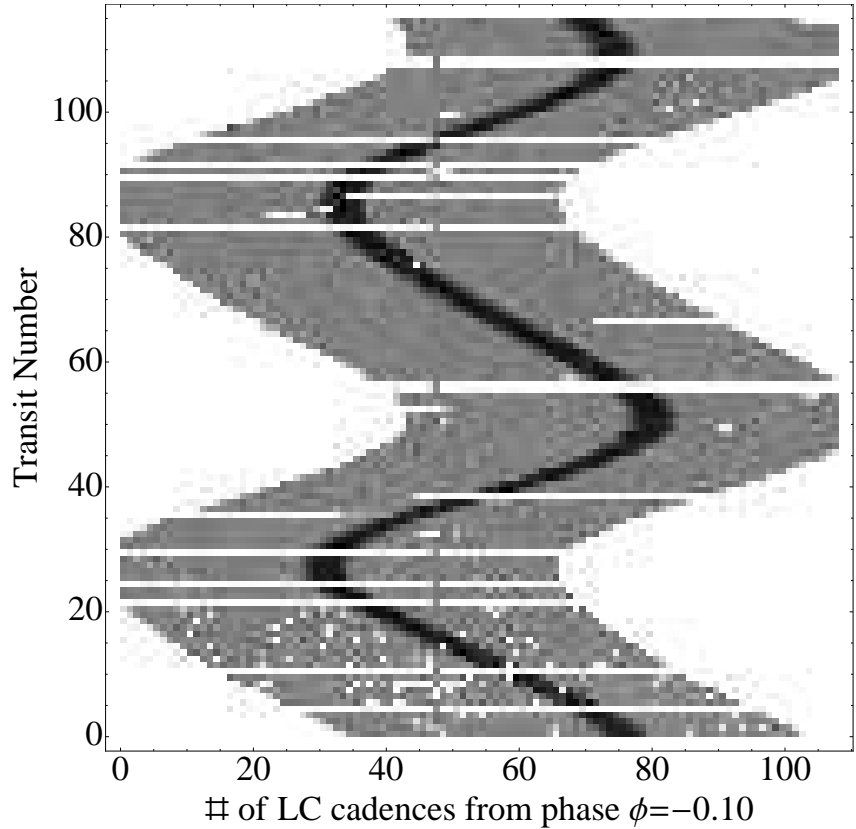}
\caption{A so-called ``river-plot'' for KOI-142.01, as used by Carter et al. 
(2012). Each row corresponds to a single transit epoch time series and the
squares within that row represent the chronological sequence of flux 
measurements represented by the color. We denote a normalized flux of unity
as gray and a normalized flux equal to a transit depth as black. White squares
denote a lack of data.}
\label{river}
\end{figure}

We downloaded the publicly available data for KOI-142 from the Mikulski
Archive for Space Telescopes (MAST), which included short-cadence (SC) data
from quarters 5-14  and long-cadence from quarters 0-14. Throughout our 
analysis, we make use of the Photometric Analysis (PA) time series. Using a
polynomial ephemeris fitted to the Mazeh et al. (2013) transit times, we 
extracted each transit epoch with $\pm0.5P_P$ worth of data either side (Fig.~\ref{river}) 
and applied the \cofiam\ (Cosine Filtering with Autocorrelation Minimization) 
detrending algorithm, which is described in detail by Kipping et al. (2013).

Although we direct those interested to Kipping et al. (2013) for details, we
briefly describe \cofiam. The algorithm is essentially a high-pass, low-cut
filter which removes periodic components with a timescale greater than that of
a pre-defined protected timescale. For this analysis, the protected timescale
was chosen to be that of three times the transit duration to ensure the transit
shape is minimally distorted by our detrending. \cofiam\ is applied to each 
transit separately, after first removing discontinuous features by eye and 
applying a moving median outlier filter. For each transit, \cofiam\ explores
typically dozens of permissible detrendings by permuting the maximum allowed
harmonic order from the maximum (corresponding to the protected timescale) to
the minimum (corresponding to the entire baseline of the time series). \cofiam\
then selects the order which minimizes the autocorrelation on a 30\,minute
timescale, as determined by the Durbin-Watson metric, $d$.

After detrending the data, the next step is to fit a transit model to the
cleaned, normalized photometry. For this purpose we use the Mandel \& Agol 
(2002) algorithm to model the transits assuming a circular orbit and the 
multimodal nested sampling algorithm \multi\ (Feroz et al. 2009, 2011) for the regression.
 Due to the presence of large dynamical variations, we wished to 
investigate the possibility of transit duration variations (TDVs) in addition to 
TTVs. This requires allowing all of the basic transit parameters to vary for 
each event and so we regress every epoch individually. In total, we detrended 
and fitted 105 transit epochs although 5 of these did not converge due to data 
gaps leading to partial transits. The $M=100$ well-fitted transits lead to $M$ 
joint-posteriors for the transit parameters. Our ``best-fit'' transit times and 
durations are computed by marginalizing each epoch's time of transit minimum 
($\tau$) and duration defined as the time it takes for the planet's center to 
cross the stellar disc to exiting under the same condition ($\tilde{T}$, see
Kipping 2010 for further details on this parameter). The final values are the 
median of each distribution and the $\pm 34.1$\% quantiles and are shown in
Figure~\ref{koi142}.

The $M$ posteriors may also be used to derive estimates of the mean transit
parameters, such as $R_P/R_*$, $a/R_*$, etc. For example, we consider a 
posterior of the mean $R_P/R_*$ made of $N$ realizations, labeled 
$i=1\rightarrow N$. For the $i^{\mathrm{th}}$ realization of the mean 
$R_P/R_*$, we calculate this as simply the sum of the $i^{\mathrm{th}}$ 
realization of each of the $M$ transit epochs divided by $M$. $N$ is limited
by the smallest length of the $M$ joint posteriors, which was $N=14034$. We
provide the median and associated $\pm34.1$\% quantiles of these mean transit
parameters in Table~\ref{tab:transit}.

Next, we investigate dynamical solutions for the large TTVs and TDVs and in
what follows we utilize the best-fit TTVs and TDVs and their associated
uncertainties.
We tested whether the measured TTVs and TDVs are consistent with gravitational perturbations from a planetary
or stellar companion of KOI-142.01, and whether a unique set of parameters can be derived to describe the 
physical and orbital properties of that companion. We examined orbits with periods between 1 day and 10 years,
including the cases of highly eccentric and/or retrograde orbits.  

The dynamical fits were obtained with a new code based on a symplectic 
$N$-body integrator known as {\tt swift\_rmvs3} (Levison \& Duncan 1994). The code computes the mid-transit 
times by interpolation. First, the transiting planet is forward propagated on the ideal Keplerian orbit starting 
from the position and velocity recorded by {\tt swift\_rmvs3} at the beginning of $N$-body time step. Second, 
the position and velocity at the end of the time step are propagated backward (again on the ideal Keplerian 
orbits). We then calculate a weighted mean of these two Keplerian trajectories such that progressively more 
(less) weight is given to the backward (forward) trajectory as the time approaches the end of the time step. 

The method described above is efficient.\footnote{The $N$-body code is nearly as fast as 
the perturbation method (Nesvorn\'y \& Morbidelli 2008, Nesvorn\'y \& Beaug\'e 2010) that was used to detect 
the non-transiting planetary companion of Kepler-46b (previously known as KOI-872b; Nesvorn\'y et al. 2012). 
Unlike the perturbation method, however, the $N$-body code can account for the resonant and near-resonant transit 
variations, which is important for KOI-142.} The required transit timing precision was achieved by setting 
the timestep to $\simeq$1/20 of the inner orbit period. The TTVs were computed relative to a linear ephemeris. 
The dynamical fits to the TTVs were obtained by minimizing
\begin{equation}
\chi_{\rm TTV}^2=\sum_{j=1}^{M} (\delta t_{{\rm O},j} - \delta t_{{\rm C},j})^2/\sigma_j^2\ , 
\end{equation}
where $M=100$ is the number of transits, $\delta t_{{\rm O},j}$ and $\delta t_{{\rm C},j}$ are the observed and 
calculated TTVs, and $\sigma_j$ is the uncertainty of $\delta t_{{\rm O},j}$.
 
The transit duration of each transit was determined from the impact parameter and the projected transit speed
at mid transit. The dynamical fits to the TDVs were obtained by minimizing
\begin{equation}
\chi_{\rm TDV}^2=\sum_{j=1}^{M} (T_{{\rm O},j} - T_{{\rm C},j})^2/\Sigma_j^2\ , 
\end{equation}
where $T_{{\rm O},j}$ and $T_{{\rm C},j}$ are the observed and calculated TDVs, and $\Sigma_j$ is the uncertainty 
of $T_{{\rm O},j}$. The simultaneous dynamical fits to the TTV and TDVs were computed by minimizing
$\chi^2 = \chi_{\rm TTV}^2 + \chi_{\rm TDV}^2$.

We tested three dynamical models: (1) coplanar orbits and circular orbit of the transiting object (model ${\cal C}$),
(2) inclined orbit of the companion and circular orbit of the transiting object (model ${\cal I}$), and (3)
a general case where the transiting object has an eccentric orbit, and the companion orbit can be inclined and 
eccentric (model ${\cal E}$). Model ${\cal C}$ has 6 parameters (two masses and four orbital parameters of the 
companion), model ${\cal I}$ has 8 parameters (six of model ${\cal C}$ plus the inclination and nodal longitude of 
companion's orbit), and model ${\cal E}$ has 10 parameters (eight of model ${\cal I}$ plus the eccentricity and 
pericenter longitude of the transiting planet). 

The remaining four parameters, namely the semimajor axis, inclination, nodal longitude and mean longitude of the 
transiting object at a given epoch, were held fixed. The semimajor axis was computed from the orbital period (obtained 
from a linear ephemeris). The inclination and nodal longitude were set from the impact parameter $b$. We used the 
transit reference system (Nesvorn\'y et al. 2012), where the nodal longitude of the transiting object is 270$^\circ$.
The mean longitude was set so that the first transit occurred at the reference epoch 
($\tau_0=2454954.62702$ BJD$_{\mathrm{UTC}}$).

\begin{figure}[t]
\epsscale{1.0}
\plotone{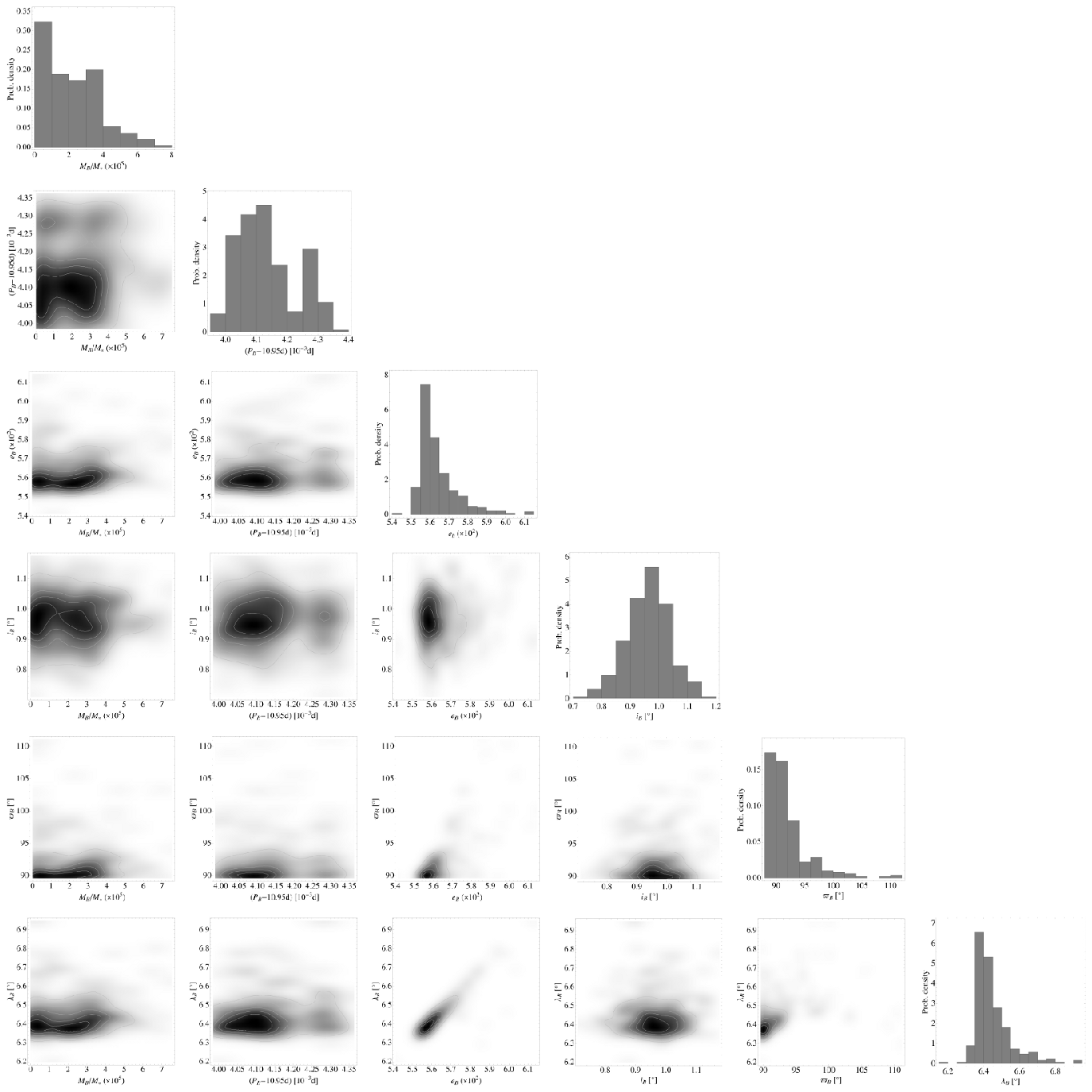}
\caption{A triangle plot for the dynamical parameters fitted to the transit
times and durations of KOI-142b. Here, the variance covariances between each
fitted parameter are easily seen.}
\label{triangleb}
\end{figure}

\begin{figure}[t]
\epsscale{1.0}
\plotone{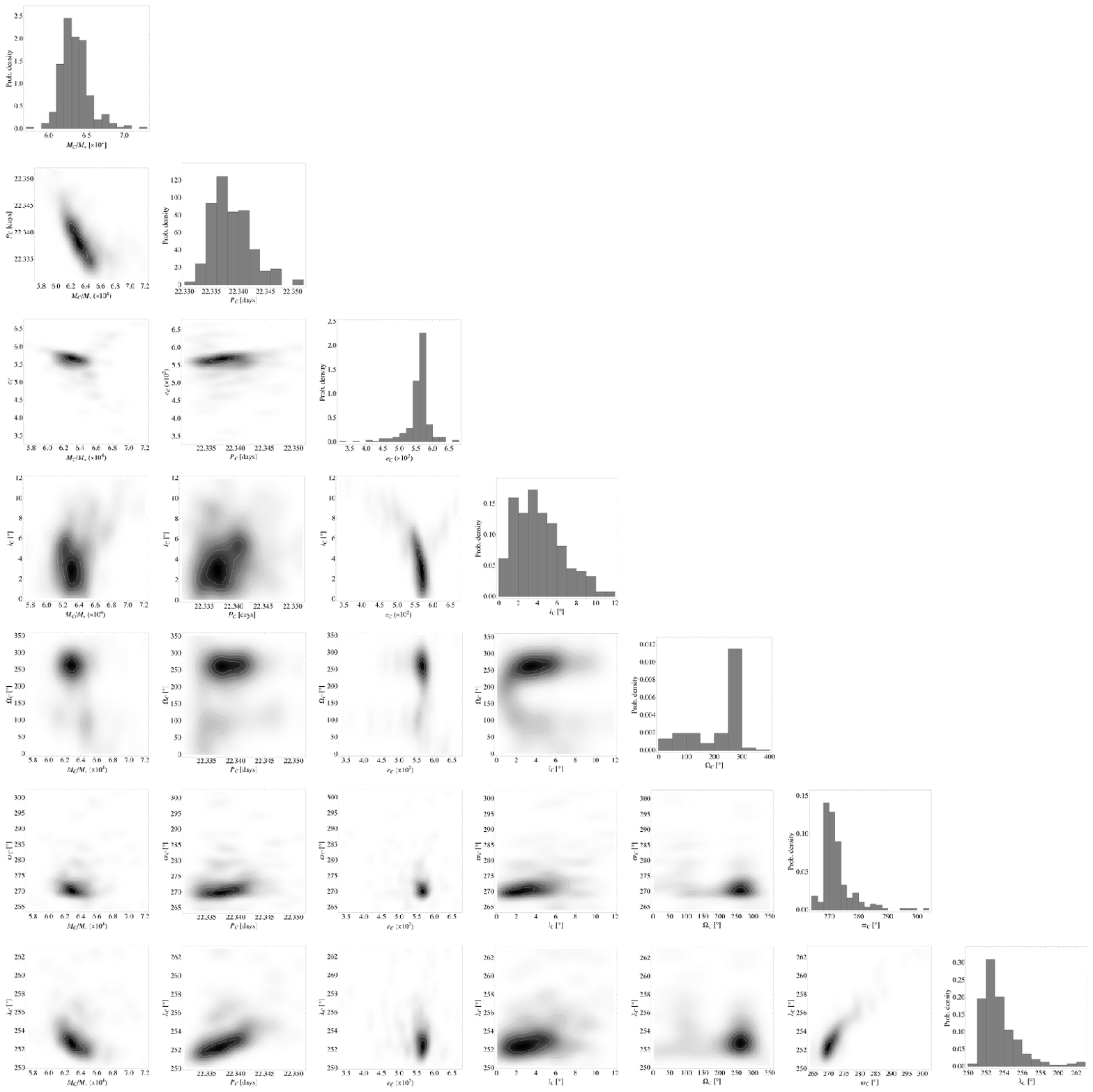}
\caption{A triangle plot for the dynamical parameters fitted to the transit
times and durations of KOI-142c. Here, the variance covariances between each
fitted parameter are easily seen.}
\label{trianglec}
\end{figure}

To compute the errors on the parameters fitted in the dynamical fits, we exploit
the posterior files derived for $\tau$ and $\tilde{T}$ for each transit epoch.
Specifically, we draw the $i^{\mathrm{th}}$ sample from the ensemble 
joint posterior for every epoch to create a fair realization of the TTV and TDV
for this $i^{\mathrm{th}}$ draw. We then repeat the process of the dynamical 
fits described above for the best-fit TTVs/TDVs. Since the draws do not 
technically have an error and the errors of the best-fit values were found to be
approximately equal anyway, we use equal weighting in the subsequent regression.
We then end up with a vector for the dynamical parameters best describing the
$i^{\mathrm{th}}$ draw of the joint posterior. This is repeated for $1000$
draws (we did not do all 14034 available realizations due to computational
constraints) allowing us to compile a joint posterior for the dynamical 
parameters (Figs. \ref{triangleb} and \ref{trianglec}). We found that 45 realizations 
resulted in unphysical results and these were discarded accordingly. The final posteriors were 
marginalized for each parameter of interest and we derived the tabulated values of 
Table~\ref{tab:dynamical} by quoting the median and the associated $\pm34.1$\%
quantiles.


Physical stellar parameters were derived by matching stellar evolution isochrones to 
the observable stellar properties. Two spectra of KOI-142 were obtained using the 
FIbre--fed \'Echelle Spectrograph (FIES) at the 2.5\,m Nordic Optical Telescope (NOT) 
at La Palma, Spain (Djupvik \& Andersen 2010) on 14 and 15 July, 2011. We used the
medium--resolution fiber with a resolving power of $\lambda/\Delta\lambda
\approx 46,\!000$ and an exposure time of 11 to 16 minutes yielding a SNR
per resolution element of 21 to 28. We used SPC (Buchhave el al. 2012) to
determine the stellar parameters of the host star, yielding an effective
temperature of $T_{\rm eff} = 5513\pm67\, {\rm K}$, a surface gravity of
$\log{g} = 4.50\pm0.12$, a metallicity of $ \rm{[m/H]}= 0.37\pm0.08$ and a
projected rotational velocity of $v \sin{i}= 1.8 \pm 0.5$ km s$^{-1}$. The
relatively low SNR spectra are at the limit of what SPC requires to extract
reliable stellar parameters, which may not be accurately reflected in the
formal uncertainties but is revealed by the relatively small percentage of
draws matching parameter sets which are allowed by stellar models  (see
below).

Rather than use $\log g$ as a
luminosity indicator, we opt to use $\rho_*$ derived from the transit light
curve. Although one might suppose we have 14034 realizations of $\rho_*$, we
actually have 14034 realizations of $\rho_{*,\mathrm{circ}}$ since we assumed
a circular orbit in the original transit fits. The true stellar density may be
easily derived using the simple correction $\rho_*=\rho_{*,\mathrm{circ}}/\Psi$
where $\Psi=(1=e\sin\omega)^3 (1-e^2)^{-3/2}$ assuming $e$ is not large (see 
Kipping 2011 for the derivation). Since we only have 955 dynamical realizations
comprising the posteriors of $e$ and $\omega$, we are only able to derive
955 realizations of $\Psi$, from which we derive 
$\Psi=1.1832_{-0.0015}^{+0.0029}$. Using the corrected 955 fair realizations of
$\rho_*$, we draw a random normal variate for $T_{\mathrm{eff}}$ and [Fe/H]
determined by SPC and then match each of these 955 draws to the Yonsei-Yale 
theoretical stellar models (Yi et al. 2001).

We find that the joint posterior of $\{\rho_*,T_{\mathrm{eff}},\mathrm{[Fe/H]}\}$ 
lies on the edge of the permissible range allowed by the stellar models, with only 
244 draws matching to a model.  The draws that match tend to have lower 
metalicity, density and temperature than the median values from the full sample 
of 955 draws.
From these trials, we marginalize over each parameter's posterior to estimate
the physical parameters for the host star KOI-142. The final stellar parameters,
defined as the median and their $\pm34.1$\% associated quantiles, are provided
in Table~\ref{tab:stellar}. We use these to compute the physical parameters of
planets b and c too, as shown in Table~\ref{tab:physical}.

\section{Results}
\begin{figure}[t]
\epsscale{0.5}
\plotone{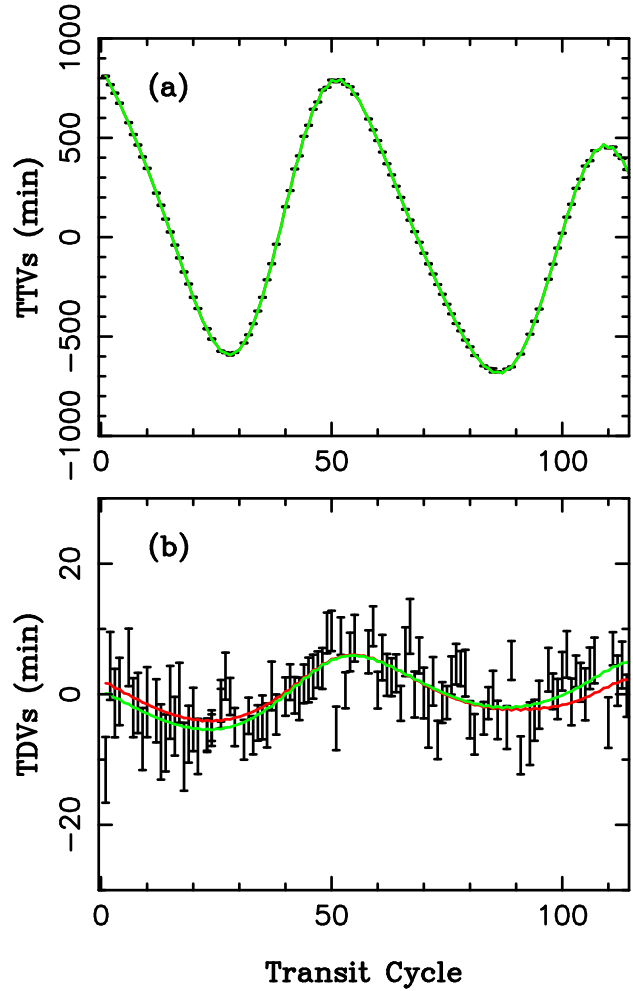}
\caption{The TTVs and TDVs of KOI-142.01. Data with associated error bars are shown in both panels.
The TTV errors are $\simeq$1-2 min and are unresolved on the scale of panel~(a). The green line
shows our best simultaneous fit to the TTVs and TDVs. The red line shows the best TTV-only fit. 
These two fits give practically identical TTVs (the two lines overlap in panel a).}
\label{koi142}
\end{figure}

KOI-142.01 shows transits with a period of $P\simeq10.95$ days. The transit timing is modulated with a TTV period of 
$P_{\rm TTV} \simeq 630$ days. The TTV amplitude exceeds 10 hours and appears to be changing with time (Fig. 
\ref{koi142}a). We detect periodic TDVs nearly in phase with the measured TTVs (Fig. \ref{koi142}b). The 
TDV amplitude is much smaller than that of the TTVs. The large $P_{\rm TTV}/P$ ratio ($\simeq$58) suggests that 
the TTVs may be related to secular, resonant or near-resonant perturbations from a companion, rather than to the 
short-periodic effects.  

\begin{figure}[t]
\epsscale{0.8}
\plotone{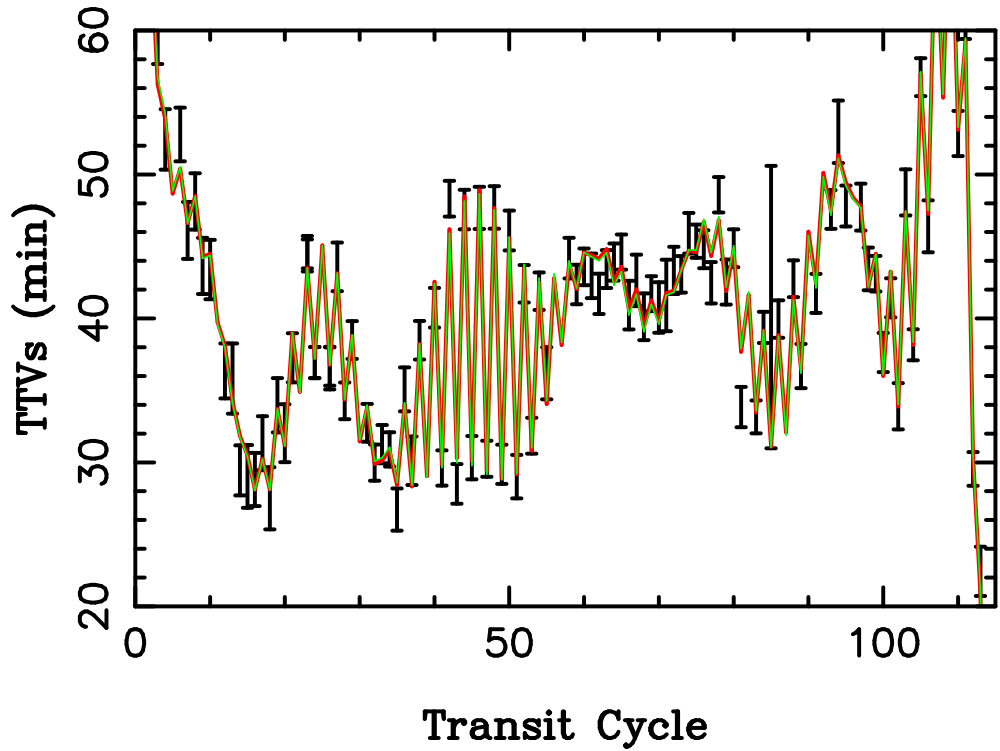}
\caption{The short-period chopping of KOI-142b's TTVs produced by the orbital conjunctions with 
KOI-142c. The main resonant terms were removed from the measured and model TTVs by a Fourier filter. 
The red and green lines show our best TTV-only and simultaneous fits from Fig. \ref{koi142}.}
\label{chop}
\end{figure}

Our detailed dynamical modeling of the measured transit variations uniquely detects a non-transiting planetary 
companion near the exterior 2:1 resonance with KOI-142.01. The uniqueness of the fits stems from the successful 
modeling of both the large near-resonant TTVs shown in Fig. \ref{koi142}a, and the short-periodic `chopping' 
produced by the orbital conjunctions between planets (Fig. \ref {chop}). The retrograde orbits near the 2:1
resonance, for example, can be ruled out because they lead to a chopping pattern that is inconsistent with 
data (conjunctions occur too often). Below we discuss the results based on the general dynamical model ${\cal E}$. 

Our best TTV-only fit gives $\chi^2_{\rm TTV}=65.7$ for 90 degrees of freedom (DOF), while the best simultaneous 
fit to the TTVs and TDVs gives $\chi^2=182$ for 190 DOF. The two fits are consistent with each other (within errors). 
They fit data very well (Fig. \ref{koi142}). All other solutions, including the highly inclined or retrograde orbits, 
can be ruled out because they give $\chi^2_{\rm TTV}>1000$. Interestingly, the TDVs expected from the TTV-only fit 
are very similar to those obtained from the simultaneous fit (Fig. \ref{koi142}b). This shows a good consistency 
of the identified solution. The best fit parameters and their errors are listed in Tables \ref{tab:dynamical} and 
\ref{tab:physical}.

We find that the companion mass is $6.3\times10^{-4}$ $M_*$, where $M_*=1.02$ M$_\odot$ (Table 3). This suggests 
a planet with mass $\simeq$0.67 M$_{\rm J}$, where M$_{\rm J}$ is the mass of 
Jupiter, or about 2.2 Saturn masses. The KOI-142.01's mass inferred from the TTVs is $\lesssim 5\times10^{-5}$ $M_*$,
implying a sub-Neptune mass planet. This is consistent with the radius ratio $R/R_*=0.039$ inferred from 
the transit analysis. With $R_*=0.96$ $R_\odot$, where $R_\odot$ is the Sun's radius, this would give 
a planetary radius similar to that of Neptune's. We therefore confirm KOI-142 as a system of two planets, 
hereafter KOI-142b and KOI-142c. 

\begin{figure}[t]
\epsscale{0.8}
\plotone{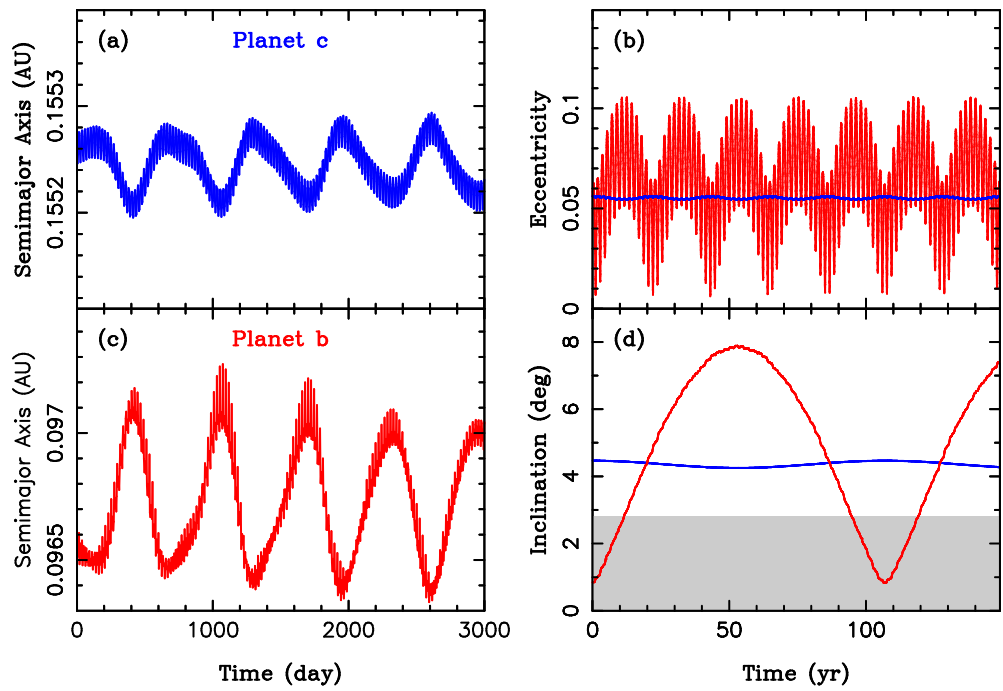}
\caption{Orbital evolution of KOI-142b and KOI-142c. The red lines denote the orbital elements of 
planet b, the blue lines denote the orbital elements of planet c. The gray area in panel (d) shows 
an approximate region where transits occur for $\Omega\simeq90^\circ$ or $\Omega\simeq 270^\circ$.}
\label{orbit}
\end{figure}

KOI-142b's and KOI-142c's orbital periods are $P_b\simeq10.95$ days and $P_c\simeq22.34$ days, respectively. The orbital 
period ratio of the two planets is therefore $P_c/P_b=2.04$ (or $\simeq2.03$ if averaged over semimajor axis 
oscillations; Fig. \ref{orbit}), indicating orbits just wide of the 2:1 mean motion resonance. This orbital configuration is 
relatively common among the KOIs and confirmed Kepler planets (Fabrycky et al. 2012). Our long-term integrations
of the system show that the orbits are stable on (at least) Gyr timescales. 

The two resonant angles, $\sigma_b=2\lambda_c-\lambda_b-\varpi_b$ and $\sigma_c=2\lambda_c-\lambda_b-\varpi_c$, where 
$\lambda$'s and $\varpi$'s are the mean and pericenter longitudes, circulate in a retrograde sense with a period of 
$P_{2:1} \simeq 630$~yr. The associated eccentricity variations of KOI-142b's orbit produce the observed TTVs and TDVs. 
A detailed interpretation of TTVs can be obtained from Lithwick et al. (2012). Here we just point out that given 
the proximity to the 2:1 resonance, $\sigma_b$ and $\sigma_c$ are not simple linear functions of time. The transit 
variations are therefore not strictly sinusoidal, as noted by Mazeh et al. (2013). 

As for the TDVs, the transit duration can be approximated by $T=\Delta/V$, where $\Delta=2R_*\sqrt{1-b^2}$ is the 
length of the transit chord, and $V = n_b a_b (1 + e_b \cos \varpi_b) / \sqrt{1-e_b^2}$ is the projected 
speed. Here, $n_b=2 \pi/P_b$ denotes the orbital frequency of KOI-142b. We find that the main contribution to the 
TDVs of KOI-142b comes from the variation of $1/(1 + e \cos \varpi)$, so that
\begin{equation}
\delta T \simeq - {2R_*\sqrt{1-b^2} \over n_b a_b} \delta k_b\ , 
\end{equation}
where $k_b = e_b \cos \varpi_b$ and $\delta k_b$ denotes the variation of $k_b$. The TDVs produced by the 
variation of $n_b a_b$ are much smaller. The vertical TDVs produced by the variation of $\sqrt{1-b^2}$ become 
important on longer timescales (see below).

Interestingly, despite their very different masses, KOI-142b and KOI-142c have similar orbital eccentricities 
(mean $e_b=0.064$ and $e_c=0.055$; Fig. \ref{orbit}). This is an interesting constraint on theories that attempt to 
explain the near-resonant planet pairs (Fabrycky et al. 2012) by tidal migration (Terquem \& Papaloizou 2007, 
Lithwick \& Wu 2012, Batygin \& Morbidelli 2012) or by planet-disk interactions (Baruteau \& Papaloizou 2013). 
The relatively large eccentricity of KOI-142c needs an explanation. 

The lines of apses of the two planets are offset so that $\Delta \varpi = \varpi_c-\varpi_b \simeq 180^\circ$ 
at the reference epoch, but $\Delta \varpi$ is not stationary, because $\varpi_b$ circulates in a prograde sense 
with the period $P_\varpi \simeq 20$ years, while the secular drift of $\varpi_c$ is much slower. The proximity 
of $\Delta \varpi$ to $180^\circ$ at the reference epoch is therefore probably coincidental. The secular 
precession of $\varpi_b$ on the $P_\varpi$ period correlates with the $\simeq$20-yr modulation of the resonant oscillations
of $e_b$ seen in Fig. \ref{orbit}b.  
 
We find a bimodal distribution of $\Omega_c$ with $\Omega_c\sim90^\circ$ or $\Omega_c\sim270^\circ$. The apparent 
absence of transits of KOI-142c implies that $i_c > 1.5^\circ$. Our best fits suggest that $2^\circ < i_c < 6^\circ$. 
While the two orbits are therefore approximately coplanar, the mutual inclination is significant, explaining 
why both planets' transits were not detected. The nodal precession period is $P_\Omega \simeq 100$ yr. The secular 
oscillations of $i_b$ happen on this timescale (Fig. \ref{orbit}b). The variations of $i_c$ have much smaller 
amplitude because $M_c/M_b \gtrsim 13$. 

\begin{figure}[t]
\epsscale{0.5}
\plotone{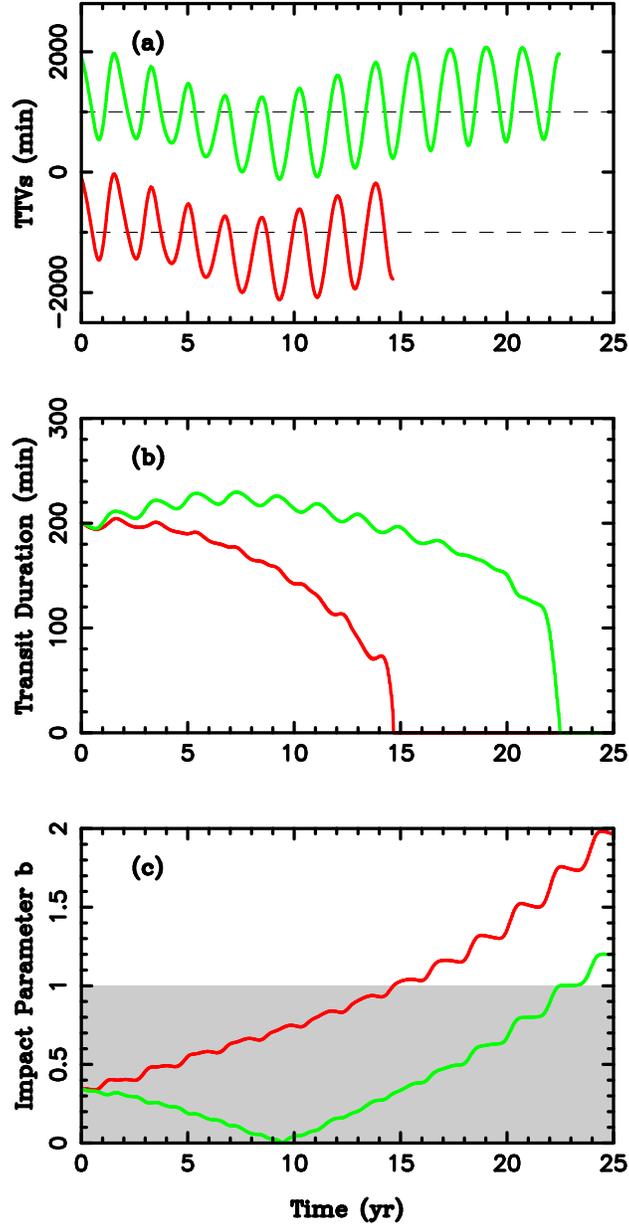}
\caption{The predicted future behavior of TTVs (panel a), transit duration (panel b), and impact 
parameter (panel c) of KOI-142b obtained from our best-fit solution. The two lines show a 
representative best solutions for $\Omega\simeq90^\circ$ (green line, $i_b=2.7^\circ$ at the reference 
epoch $t=0$) and $\Omega\simeq270^\circ$ (red line, $i_b=4.5^\circ$ at $t=0$). The TTVs in panel
(a) were offset by $\pm1000$ min for clarity. The gray area in panel (c) shows the region where 
transits occur ($b<1$).}
\label{predict}
\end{figure}

The secular dynamics of KOI-142b has interesting implications for the future behavior of TTVs and TDVs. As shown in 
Fig. \ref{predict}a, the TTV signal is expected to be modulated on the $P_\varpi$ period. Moreover, the secular 
changes of $i_b$ and $\Omega_b$ will profoundly affect the transit duration (Fig. \ref{predict}b) and the impact 
parameter (Fig. \ref{predict}b). Depending on whether $\Omega_c\sim90^\circ$ or $\Omega_c\sim270^\circ$ at the present,
the transit impact parameter $b_b$ is expected to decrease (for $\Omega_c\simeq90^\circ$) or increase 
(for $\Omega_c\simeq270^\circ$) over the next few years. Monitoring $b_b$ over the next decade will thus help to 
remove the $\Omega_c$ degeneracy, and identify the correct value of $\Omega_c$.

In any case, we predict that KOI-142b should stop transiting within the next 10-25~years. The reason for that is
the the orbital plane KOI-142b is inclined relative to the more massive KOI-142c, must precess around KOI-142c's 
orbital plane. As it precesses, $i_b$ relative to the transit plane will increase to the point when $b_b>1$. The 
exact time interval during which the transits of KOI-142b will still occur depends on 
the present $i_c$ and $\Omega_c$. The transit window can be as long as several decades for $\Omega_c\simeq90^\circ$ 
and the smallest acceptable values of $i_c$ ($\simeq$1.5$^\circ$-2$^\circ$). On the other hand, in can be as short
as $\simeq$10 years if $\Omega_c\simeq270^\circ$ and $i_c\sim5^\circ$. 

\begin{figure}[t]
\epsscale{0.5}
\plotone{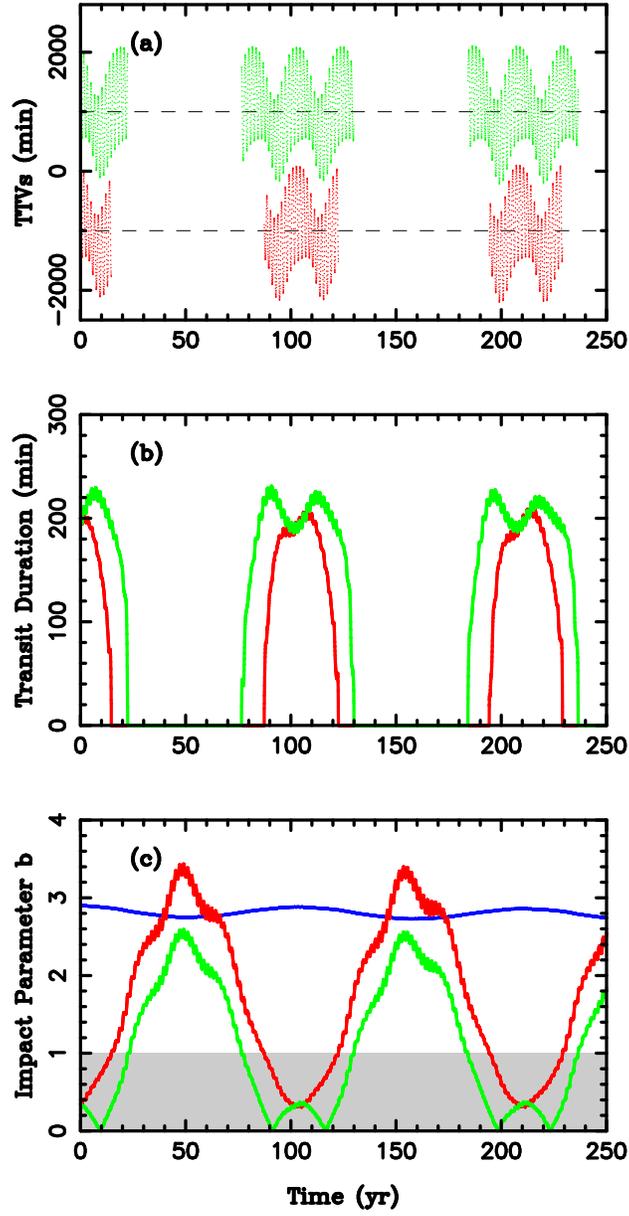}
\caption{The same as Fig. \ref{predict} but for an extended time interval. The blue line in panel (c)
shows the impact parameter of KOI-142c corresponding to the best fit solution shown in Fig. \ref{orbit}d.}
\label{predict2}
\end{figure}

The transit windows of KOI-142b should last 30-50 years, and should re-appear with the periodicity of $P_\Omega\simeq100$ 
yr (Fig. \ref{predict2}). It is unlikely that the transits of KOI-142c could ever be observed from the Earth, because 
the orbital plane of KOI-142c remains inclined with respect to the transit plane (assuming that there are no 
additional massive bodies in the system). We find that $1.5<b_c<4$ with the best fit values occurring for
$b_c \simeq 3$ (Fig. \ref{predict2}). 

We predict that the RV measurements of KOI-142 should reveal at least two basic periods. The RV term with a 
22.34\,d period, corresponding to KOI-142c, should have $K \simeq 48$~m~s$^{-1}$ amplitude (Table 4). The 
amplitude of the 10.95\,d period term, corresponding to KOI-142b, is uncertain because we only have an upper limit 
on $M_b$. With $M_b/M_*=5\times 10^{-5}$, we obtain $K \simeq 5$~m~s$^{-1}$.
\section{Summary}
We performed a detailed analysis of the KOI-142 transit photometry. The observed TTVs with an amplitude over 
10 hours are some of the largest ever detected for a KOI. We found evidence for periodic TDVs that are nearly in 
phase with the measured TTVs, but are much smaller in amplitude ($\simeq$5 min). To our knowledge, this is the first 
time that TDVs were unequivocally detected for an exoplanet. A thorough dynamical analysis was then conducted to 
find plausible interpretations for the measured TTVs and TDVs.  

Based on this analysis, we confirm KOI-142 as a system of two planets just wide of the 2:1 resonance (average 
$P_c/P_b=2.03$). The inner Neptune-class planet, KOI-142b, shows transits at the present epoch. We predict 
that these transits should disappear in the next few decades. Monitoring the impact parameter (or equivalently 
the transit duration) of KOI-142b will help to better constrain the inclination and nodal longitude of
KOI-142c. The outer sub-Jovian mass planet, KOI-142c, is currently not transiting, and we find that the viewing 
geometry and dynamics preclude the visibility of KOI-142c transits from Earth in the future.

The dynamical configuration of KOI-142b's and KOI-142c's orbits near the 2:1 resonance can be used to constrain
the formation and migration history of the two planets (Beaug\'e et al., in preparation). The relatively large 
orbital eccentricity of KOI-142c (mean $e_c=0.055$) cannot be explained by present gravitational perturbations 
from KOI-142b, or by tidal migration (Lithwick \& Wu 2012, Batygin \& Morbidelli 2012). It was probably established 
early, possibly during KOI-142c's formation in the proto-planetary gas disk (e.g., Lega et al. 2013). 

\acknowledgments
We thank the Kepler Science Team, especially the DAWG, for making the data used here available.
DN acknowledges support from NSF AST-1008890. 
DMK is supported by the NASA Sagan fellowship.
JH acknowledges support from NSF AST-1108686.
GB acknowledges support from NASA grant NNX09AB29G.

\begin{table}
\caption{\emph{Mean transit parameters derived by taking the average
of each posterior sample across the $M=100$ joint posteriors derived for
each transit epoch.}}
\vspace*{2.mm}
\centering 
\begin{tabular}{c c} 
\hline\hline 
Parameter & Value \\ [0.5ex]
\hline 
$P_b$\,[days] & $10.954204_{-0.00143}^{+0.000064}$ \\
$R_b/R_*$ & $0.0390_{-0.0037}^{+0.0041}$ \\
$a_b/R_*$\,(circ) & $24.8_{-1.4}^{+1.3}$ \\
$b_b$ & $0.372_{-0.029}^{+0.028}$ \\
$i_b$\,[$^{\circ}$] & $89.140_{-0.089}^{+0.085}$ \\
$\rho_{*,b}$\,(circ)\,[g\,cm$^{-3}$] & $2.40_{-0.39}^{+0.41}$ \\
$\tilde{T}_b$\,[hours] & $3.13_{-0.16}^{+0.19}$ \\
$T_{14,b}$\,[hours] & $3.28_{-0.17}^{+0.20}$ \\
$T_{23,b}$\,[hours] & $2.99_{-0.15}^{+0.19}$ \\
$T_{12,b} \simeq T_{34,b}$\,[hours] & $0.142_{-0.014}^{+0.017}$ \\ [1ex]
\hline\hline 
\end{tabular}
\label{tab:transit} 
\end{table}

\begin{table}
\caption{\emph{Final values of the free parameters used in the dynamical fits
of the transit times and durations (TTVs and TDVs). The upper limit of 
$M_b/M_*$ corresponds to the 95\% quantile of the posterior distribution. The orbital 
inclination $i$ and nodal longitude $\Omega$ are given with respect to the transit reference 
system (Nesvorn\'y et al. 2012).}}
\vspace*{2.mm}
\centering 
\begin{tabular}{c c c} 
\hline\hline 
Parameter & Planet b & Planet c \\ [0.5ex]
\hline 
$M_P/M_*$ & $<5.2\times 10^{-5}$ & $(6.32_{-0.13}^{+0.19})\times 10^{-4}$ \\
$P_P$\,[days] & $10.954204_{-0.00143}^{+0.000064}$ & $22.3383_{-0.0031}^{+0.0036}$ \\
$i$\,[$^{\circ}$] & $0.931_{-0.079}^{+0.088}$ & $3.8_{-2.1}^{+2.4}$ \\
$e$ & $0.05608_{-0.00047}^{+0.00108}$ & $0.0564_{-0.0023}^{+0.0015}$ \\
$\varpi$\,[$^{\circ}$] & $90.9_{-1.1}^{+3.7}$ & $271.1_{-1.8}^{+4.8}$\\
$\Omega$\,[$^{\circ}$] & $270$ & $259_{-164}^{+11}$ \\
$\lambda$\,[$^{\circ}$] & $6.419_{-0.051}^{+0.094}$ & $253.0_{-1.0}^{+2.1}$ \\ [1ex]
\hline\hline 
\end{tabular}
\label{tab:dynamical} 
\end{table}

\begin{table}
\caption{\emph{Derived stellar parameters found by matching isochrones to
the SPC-derived stellar effective temperature and metalicity as well as the
light curve derived stellar density (accounting for orbital eccentricity).
For comparison, we show the Kepler Input Catalog (KIC) parameters derived by
Brown et al. (2012).}}
\vspace*{2.mm}
\centering 
\begin{tabular}{c c c} 
\hline\hline 
Parameter & KIC & This work \\ [0.5ex]
\hline
\emph{Fitted param.} & & \\ 
\hline 
$M_*$\,[$M_{\odot}$] & 0.96 & $1.022_{-0.026}^{+0.023}$ \\
$R_*$\,[$R_{\odot}$] & 0.74 & $0.961_{-0.024}^{+0.020}$ \\
$T_{\mathrm{eff}}$\,[K] & $5361 \pm 135$ & $5513 \pm 67$ \\
$[{\rm Fe/H}]$ \, [dex] & $-0.10 \pm 0.20$ & $+0.37 \pm 0.08$ \\
$\log g$\,[dex] & $4.68 \pm 0.25$ & $4.482_{-0.016}^{+0.018}$ \\
$L_*$\,[$R_{\odot}$] & - & $0.755_{-0.067}^{+0.056}$ \\
Age\,[Gyr] & - & $2.45_{-0.77}^{+1.20}$ \\
$M_V$ & - & $5.183_{-0.090}^{+0.116}$ \\
Distance\,[pc] &  - & $385_{-20}^{+16}$ \\ [1ex]
\hline\hline 
\end{tabular}
\label{tab:stellar} 
\end{table}

\begin{table}
\caption{\emph{Physical planetary parameters derived by combining the stellar
parameters shown in Table~\ref{tab:stellar} and the dynamical parameters shown
in Table~\ref{tab:dynamical}. The upper limits correspond to the 95\% 
quantile of the posterior distribution. The orbital inclination $i$ and 
nodal longitude $\Omega$ are given with respect to the transit reference system 
(Nesvorn\'y et al. 2012). }}
\vspace*{2.mm}
\centering 
\begin{tabular}{c c c} 
\hline\hline 
Parameter & Planet b & Planet c \\ [0.5ex]
\hline 
$M_P$ [$M_{\oplus}]$ & $<17.6$ & $215.9_{-7.5}^{+7.6}$ \\
$R_P$ [$R_{\oplus}]$ & $4.23_{-0.39}^{+0.30}$ & - \\
$\rho_P$ [g\,cm$^{-3}$] & $0.48_{-0.46}^{+0.54}$ & - \\
$P_P$\,[days] & $10.954204_{-0.00143}^{+0.000064}$ & $22.3383_{-0.0031}^{+0.0036}$ \\
$i$\,[$^{\circ}$] & $0.962_{-0.078}^{+0.077}$ & $3.7_{-2.1}^{+2.7}$ \\
$e$ & $0.05611_{-0.00052}^{+0.00121}$ & $0.05628_{-0.0028}^{+0.0014}$ \\
$\varpi$\,[$^{\circ}$] & $90.7_{-1.0}^{+3.4}$ & $271.1_{-2.0}^{+4.7}$\\
$\Omega$\,[$^{\circ}$] & $270$ & $259_{-164}^{+11}$ \\
$\lambda$\,[$^{\circ}$] & $6.425_{-0.059}^{+0.098}$ & $253.0_{-1.1}^{+2.3}$ \\
$K$\,[m\,s$^{-1}$] & $<5.0$ & $48.2_{-1.3}^{+1.3}$ \\ [1ex]
\hline\hline 
\end{tabular}
\label{tab:physical} 
\end{table}

\clearpage

\clearpage

\clearpage

\clearpage

\clearpage

\clearpage

\clearpage

\clearpage


\begin{thebibliography}

\bibitem[Agol et al.(2005)]{2005MNRAS.359..567A} Agol, E., Steffen, J., 
Sari, R., \& Clarkson, W.\ 2005, \mnras, 359, 567 

\bibitem[Baruteau 
\& Papaloizou(2013)]{2013arXiv1301.0779B} Baruteau, C., \& Papaloizou, J.~C.~B.\ 2013, arXiv:1301.0779 

\bibitem[Batygin 
\& Morbidelli(2013)]{2013AJ....145....1B} Batygin, K., \& Morbidelli, A.\ 2013, \aj, 145, 1 

\bibitem[Buchhave et al.(2012)]{2012yCatp038048601B} Buchhave, L.~A., 
Latham, D.~W., Johansen, A., et al.\ 2012, VizieR Online Data Catalog, 380, 
48601 

\bibitem[Djupvik \& Andersen(2010)]{djupvik:2010} Djupvik, A.~A., \&
  Andersen, J.\ 2010, in ``Highlights of Spanish Astrophysics V''
  eds. J.~M.~Diego, L.~J.~Goicoechea, J.~I.~Gonz\'alez-Serrano, \&
  J.~Gorgas (Springer: Berlin), p. 211

\bibitem[Fabrycky et al.(2012)]{2012arXiv1202.6328F} Fabrycky, D.~C., 
Lissauer, J.~J., Ragozzine, D., et al.\ 2012, arXiv:1202.6328 

\bibitem[Feroz et al.(2009)]{2009MNRAS.398.1601F} Feroz, F., Hobson, M.~P., 
\& Bridges, M.\ 2009, \mnras, 398, 1601 

\bibitem[Feroz et al.(2011)]{2011ascl.soft09006F} Feroz, F., Hobson, M.~P., 
\& Bridges, M.\ 2011, Astrophysics Source Code Library, 9006 

\bibitem[Ford et al.(2011)]{2011ApJS..197....2F} Ford, E.~B., Rowe, J.~F., 
Fabrycky, D.~C., et al.\ 2011, \apjs, 197, 2 

\bibitem[Ford et al.(2012)]{2012ApJ...756..185F} Ford, E.~B., Ragozzine, 
D., Rowe, J.~F., et al.\ 2012, \apj, 756, 185 

\bibitem[Holman 
\& Murray(2005)]{2005Sci...307.1288H} Holman, M.~J., \& Murray, N.~W.\ 2005, Science, 307, 1288 

\bibitem[Holman et al.(2010)]{2010Sci...330...51H} Holman, M.~J., Fabrycky, 
D.~C., Ragozzine, D., et al.\ 2010, Science, 330, 51 

\bibitem[Kipping(2010)]{2010MNRAS.407..301K} Kipping, D.~M.\ 2010, \mnras, 
407, 301 

\bibitem[Kipping(2011)]{2011MNRAS.416..689K} Kipping, D.~M.\ 2011, \mnras, 
416, 689 

\bibitem[Kipping et al.(2012)]{2012ApJ...750..115K} Kipping, D.~M., Bakos, 
G.~{\'A}., Buchhave, L., Nesvorn{\'y}, D., 
\& Schmitt, A.\ 2012, \apj, 750, 115 

\bibitem[Kipping et al.(2013)]{2013arXiv1301.1853K} Kipping, D.~M., 
Hartman, J., Buchhave, L.~A., et al.\ 2013, arXiv:1301.1853 

\bibitem[a]{lega} Lega, A., Morbidelli, A. Nesvorn\'y, D., MNRAS, in press

\bibitem[Levison 
\& Duncan(1994)]{1994Icar..108...18L} Levison, H.~F., \& Duncan, M.~J.\ 1994, Icarus, 108, 18 

\bibitem[Lissauer et al.(2011)]{2011ApJS..197....8L} Lissauer, J.~J., 
Ragozzine, D., Fabrycky, D.~C., et al.\ 2011a, \apjs, 197, 8 

\bibitem[Lissauer et al.(2011)]{2011Natur.470...53L} Lissauer, J.~J., 
Fabrycky, D.~C., Ford, E.~B., et al.\ 2011b, \nat, 470, 53 

\bibitem[Lithwick 
\& Wu(2012)]{2012ApJ...756L..11L} Lithwick, Y., \& Wu, Y.\ 2012, \apjl, 756, L11 

\bibitem[Lithwick et al.(2012)]{2012ApJ...761..122L} Lithwick, Y., Xie, J., 
\& Wu, Y.\ 2012, \apj, 761, 122 

\bibitem[Mandel 
\& Agol(2002)]{2002ApJ...580L.171M} Mandel, K., \& Agol, E.\ 2002, \apjl, 580, L171 

\bibitem[Mazeh et al.(2013)]{2013arXiv1301.5499M} Mazeh, T., Nachmani, G., 
Holczer, T., et al.\ 2013, arXiv:1301.5499 

\bibitem[Miralda-Escud{\'e}(2002)]{2002ApJ...564.1019M} Miralda-Escud{\'e}, 
J.\ 2002, \apj, 564, 1019 

\bibitem[Nesvorn{\'y} 
\& Morbidelli(2008)]{2008ApJ...688..636N} Nesvorn{\'y}, D., \& Morbidelli, A.\ 2008, \apj, 688, 636 

\bibitem[Nesvorn{\'y} 
\& Beaug{\'e}(2010)]{2010ApJ...709L..44N} Nesvorn{\'y}, D., \& Beaug{\'e}, C.\ 2010, \apjl, 709, L44 

\bibitem[Nesvorn{\'y} et al.(2012)]{2012Sci...336.1133N} Nesvorn{\'y}, D., 
Kipping, D.~M., Buchhave, L.~A., et al.\ 2012, Science, 336, 1133 

\bibitem[Steffen et al.(2012)]{2012ApJ...756..186S} Steffen, J.~H., Ford, 
E.~B., Rowe, J.~F., et al.\ 2012, \apj, 756, 186 

\bibitem[Terquem 
\& Papaloizou(2007)]{2007ApJ...654.1110T} Terquem, C., \& Papaloizou, J.~C.~B.\ 2007, \apj, 654, 1110 


\bibitem[Yi et al.(2001)]{2001ApJS..136..417Y} Yi, S., Demarque, P., Kim, 
Y.-C., et al.\ 2001, \apjs, 136, 417 

\end{thebibliography}
\end{document}